\documentclass[sigconf]{acmart}
\usepackage{multirow}
\usepackage{amsmath,bm}
\usepackage{booktabs}
\usepackage{pifont}
\usepackage{booktabs, caption, makecell}

\usepackage[shortlabels]{enumitem}
\usepackage{algorithm}
\usepackage{algorithmic}
\AtBeginDocument{%
  \providecommand\BibTeX{{%
    \normalfont B\kern-0.5em{\scshape i\kern-0.25em b}\kern-0.8em\TeX}}}

\copyrightyear{2023}
\acmYear{2023}
\setcopyright{acmlicensed}\acmConference[CIKM '23]{Proceedings of the 32nd ACM International Conference on Information and Knowledge Management}{October 21--25, 2023}{Birmingham, United Kingdom}
\acmBooktitle{Proceedings of the 32nd ACM International Conference on Information and Knowledge Management (CIKM '23), October 21--25, 2023, Birmingham, United Kingdom}
\acmPrice{15.00}
\acmDOI{10.1145/3583780.3614965}
\acmISBN{979-8-4007-0124-5/23/10}

\begin{document}

\title{Meta-Learning with Adaptive Weighted Loss\\for Imbalanced Cold-Start Recommendation}
\renewcommand{\shorttitle}{Meta-Learning with Adaptive Weighted Loss for Imbalanced Cold-Start Recommendation}

\author{Minchang Kim}
\authornote{Both authors contributed equally to this research.}
\affiliation{%
  \institution{Seoul National University}
  \city{Seoul}
  \country{Republic of Korea}
}
\email{minchang.kim@snu.ac.kr}

\author{Yongjin Yang}
\authornotemark[1]
\affiliation{%
  \institution{Seoul National University}
  \city{Seoul}
  \country{Republic of Korea}
}
\email{dyyjkd@snu.ac.kr}

\author{Jung Hyun Ryu}
\affiliation{%
  \institution{Seoul National University}
  \city{Seoul}
  \country{Republic of Korea}
}
\email{jhryu30@snu.ac.kr}

\author{Taesup Kim}
\authornote{Corresponding author}
\affiliation{%
  \institution{Seoul National University}
  \city{Seoul}
  \country{Republic of Korea}
}
\email{taesup.kim@snu.ac.kr}

\renewcommand{\shortauthors}{Minchang Kim, Yongjin Yang, Jung Hyun Ryu, and Taesup Kim}

\begin{abstract}
Sequential recommenders have made great strides in capturing a user's preferences. Nevertheless, the cold-start recommendation remains a fundamental challenge as they typically involve limited user-item interactions for personalization. Recently, gradient-based meta-learning approaches have emerged in the sequential recommendation field due to their fast adaptation and easy-to-integrate abilities. The meta-learning algorithms formulate the cold-start recommendation as a few-shot learning problem, where each user is represented as a task to be adapted. While meta-learning algorithms generally assume that task-wise samples are evenly distributed over classes or values, user-item interactions in real-world applications do not conform to such a distribution (e.g., watching favorite videos multiple times, leaving only positive ratings without any negative ones). Consequently, imbalanced user feedback, which accounts for the majority of task training data, may dominate the user adaptation process and prevent meta-learning algorithms from learning meaningful meta-knowledge for personalized recommendations. To alleviate this limitation, we propose a novel sequential recommendation framework based on gradient-based meta-learning that captures the imbalanced rating distribution of each user and computes adaptive loss for user-specific learning. Our work is the first to tackle the impact of imbalanced ratings in cold-start sequential recommendation scenarios. Through extensive experiments conducted on real-world datasets, we demonstrate the effectiveness of our framework.
\end{abstract}



\maketitle

\section{Introduction}
When modeling a user's dynamic preferences, temporal user-item information plays an important role. For this reason, sequential recommender systems have continuously evolved in the field of recommendation systems~\cite{fang2020deep,wang2019sequential}. Sequence models leverage temporal information and excel at compressing the user's historical behaviors to a single representation~\cite{li2017neural}. However, since sequential recommender systems mainly rely on the total number of historical interactions, they are fragile to the cold-start problem, where user-item interactions are too short that only limited user information is available~\cite{zheng2021cold}.

The user cold-start recommendation is both inevitable and crucial in a competitive business environment. When new users come to an online service (e.g., e-commerce, streaming services), it is a significant moment that determines the sustainability of a business. Users often decide whether or not to continue using the service based on their initial interactions. If they are dissatisfied with the quality of service from the onset, they might easily switch to an alternative service. While businesses strive to gather user information promptly to offer better choices, increasing concerns about privacy mean that many users are hesitant to share their personal details.

Recently, various studies have widely adopted gradient-based meta-learning algorithms, such as model-agnostic meta-learning (MAML)~\cite{finn2017model}, in recommendation systems. It has the advantage of being smoothly incorporated with existing recommender networks optimized by gradient descent. The basic idea behind meta-learning in recommendation systems is to formulate the user cold-start recommendation as a few-shot learning problem and train the model to adapt to new (unseen) users rapidly. MeLU~\cite{lee2019melu} integrates user and item attributes with the MAML algorithm, and MAMO~\cite{dong2020mamo} further uses task-specific and feature-specific memories to keep individual users' preferences. MetaCSR~\cite{huang2022learning} incorporates the gradient-based meta-learner and the diffusion representer that is composed of graph convolutional networks (GCNs)~\cite{kipf2016semi} to model high-order user-item correlation without side information (e.g., age, job). So far, various meta-learning approaches have effectively improved the cold-start recommendation performance.

\begin{figure}[t]
    \centering
    \includegraphics[width=0.9\columnwidth]{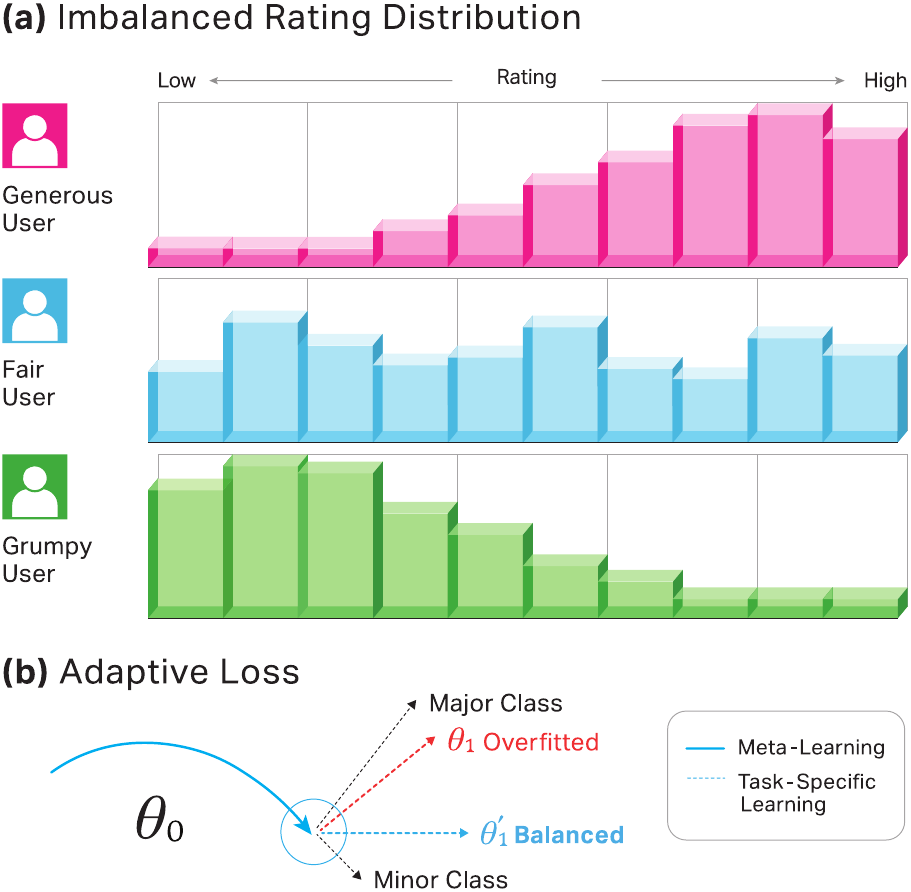}
    \caption{The concept of our method. (a) Individual users have various rating distributions. The mean,  variance, and skewness are different in each case. (b) Adaptive weighted loss corrects inner-loop adaptation path for each user. We expect minor classes could be considered more for task-specific learning to prevent overfitting.}
    \label{fig:1}
\end{figure}

In a general setting, the MAML algorithm supposes that the number of examples per class is fixed (i.e., $N$-way $k$-shot setting). However, as illustrated in Figure~\ref{fig:1}, in real-world scenarios, each user possesses individual criteria for rating items. Consequently, user ratings are not uniformly distributed. This discrepancy can lead to over- and under-fitting issues, which prevent the model from learning meaningful meta-knowledge~\cite{lee2019learning}.

Traditionally, studies on sequential recommender systems have primarily concentrated on users' implicit feedback (e.g., clicks, views) rather than explicit feedback (e.g., ratings, likes)~\cite{hu2008collaborative}. As a result, the negative impact of imbalanced rating distributions from explicit feedback on sequential recommendation performance has been underrepresented. In this work, we aim for a deeper understanding of a user's preferences and better reflection of real-world data by simultaneously utilizing both implicit and explicit feedback (multi-behavior data). Several research efforts have already been undertaken from a multi-behavior data perspective~\cite{gao2019neural,jin2020multi}.

We propose a novel {\bf ME}ta-learning based sequential recommender with adaptive weighted {\bf LO}ss for imbalanced cold-start recommendation ({\bf MELO}). Specifically, we have improved existing meta-learning based sequential recommendation methods for the user cold-start problem by handling the rating imbalance problem with our proposed adaptive loss. Our framework consists of (1) the sequential recommender modeling temporal dynamics in user-item interactions, (2) the gradient-based meta-learner that makes the recommender quickly adapt to individual users by observing only a few interactions, and (3) the adaptive weighted loss, which dynamically adjusts the importance of each interaction during the user adaptation process by capturing the task state.

The main contribution of this work is summarized as follows: 
\begin{itemize}
\item It is the first work to tackle the negative impact of imbalanced rating distribution in real-world cold-start recommendation scenarios. 
\item We propose the adaptive weighted loss that recognizes the rating imbalance based on temporal dynamics and corrects the learning path during the meta-learning process.
\item Our framework can be easily integrated with any existing sequential recommendation models, such as GRU4Rec, NARM, SASRec, and BERT4Rec.
\item We extensively validate our framework with existing sequential recommenders on real-world datasets to demonstrate the effectiveness of our proposal.
\end{itemize}

\section{Related Work}
\subsection{Sequential Recommendation}
Over the past few years, state-of-the-art sequential recommender systems have tried to model user-item interactions and focus on understanding a users' preferences \cite{wang2019sequential}. GRU4Rec~\cite{hidasi2015session} learns the contextual relationship between users and items in the sequence through GRU layers. NARM~\cite{li2017neural} similarly incorporates an attention mechanism into GRU architectures. SASRec~\cite{kang2018self} brings transformer architectures to build higher-capacity models. BERT4Rec~\cite{sun2019bert4rec} employs a bidirectional encoder and masked training techniques to improve unidirectional transformer models. 

Recommendation systems typically take two types of user feedback data: implicit and explicit \cite{he2017neural}. Implicit feedback is information obtained indirectly by monitoring users' behavior, such as item clicks and page views. On the other hand, explicit feedback is a response obtained directly from users according to their preferences for a product, such as ratings and likes. Sequential recommender systems excel at modeling users' indirect behavior, and most of sequential recommender researches have targeted implicit feedback only. However, there's a growing body of recent studies that utilize multi-behavior data, incorporating both implicit and explicit feedback, to achieve a more comprehensive understanding of users \cite{gao2019neural,jin2020multi}. In this context,  we propose a novel method to mitigate the imbalanced distribution of explicit user ratings while modeling the temporal dynamics of implicit user-item interaction.

\subsection{Meta-Learning for Cold-Start Problem}
The core idea of recommendation systems with gradient-based meta-learning is to learn the meta-knowledge that initializes the model parameter for personalized recommendation. MeLU~\cite{lee2019melu} leverages side information with the MAML algorithm~\cite{finn2017model}. They locally update the decision-making layer based on each user's pattern and globally update user-item embeddings and whole layers. MAMO~\cite{dong2020mamo} designs two memory networks and guides the global sharing initialization parameter not to fall into the local optima of irregular users. The feature-specific memory helps to initialize the recommender model, and the task-specific memory guides the prediction process. MetaCSR~\cite{huang2022learning} incorporates high-order embeddings of user-item generated by graph neural networks and the gradient-based meta-learning algorithm to extract sharable knowledge of prior users. Mecos~\cite{zheng2021cold} encodes sequence pairs and trains the matching network to connect the target item to potential users. They present a practical approach for the item cold-start problem.

Recent works propose various meta-learning based approaches for the cold-start recommendation. However, most approaches do not seriously consider the imbalance of rating distribution, which easily varies from one user to another during the inner-loop optimization process. In this work, we highlight the importance of recognizing the imbalanced rating distribution and propose the adaptive weighted loss to accordingly and dynamically calibrate the learning process.

\subsection{Meta-Learning for Imbalanced Problem}
The imbalance data generally degrades the learning quality, and research on meta-learning for the imbalanced problem is attracting attention. MeTAL~\cite{baik2021meta} proposes a learnable approach for training a loss function, aiming for enhanced generalization across various applications. The loss function network computes task-adaptive loss with embeddings generated by the task state network. Both the loss function and task state neural networks are optimized each time during an outer-loop update. Bayesian TAML~\cite{lee2019learning} introduces three learnable variables to scale the loss function. They process statistical information such as the mean, variance, and cardinality for the input of balancing variable networks. They also design the variational inference framework to alleviate the randomness of statistical values. PALM~\cite{yu2021personalized} proposes a top-K nearest neighbor search to obtain different learning rates for individual users. A tree-based method and a memory-agnostic regularizer are employed to increase the efficiency of search and storage operations. ATS~\cite{yao2021meta} introduces a neural task scheduler that identifies informative tasks (excluding noisy labels) for the task adaptation process. They suggest two input features, which represent the characteristics of a candidate task, to avoid the effort of finding the optimal combination heuristically. 

Overall, the above studies propose various learnable ways to adjust the loss values to train the model correctly. However, the proposed methods either use complex networks (e.g., many inputs and parameters) or expensive algorithms (e.g., search and sort) to compute. In this work, we suggest a more straightforward and automated way to adjust the loss function by sample-wise adaptive weights computed in a sequence manner.

\section{Preliminary}
\subsection{Problem Formulation}
The sequential recommendation task assumes $\smash{\mathcal{U} = \{u_1, u_2, \dots, u_{n}\}}$ as a set of users and $\smash{\mathcal{V} = \{v_1, v_2, \dots,v_{m}\}}$ as a set of items, where $n$ and $m$ are respectively the number of users and items. A sequence of user-item interactions generated by user $\smash{u_i \in \mathcal{U}}$ is represented as $\smash{\mathrm{X}_i = \{v_{1}^{i}, v_{2}^{i}, \dots, v_{t}^{i}\}}$ in chronological order at time step $\mathnormal{t}$. Here, $\mathnormal{t}$ denotes the interaction order rather than the absolute timestamp, similar to previous works \cite{wu2017recurrent, kang2018self}. In a rating prediction task, let $\smash{\mathrm{Y}_i = \{y_{1}^{i}, y_{2}^{i}, \cdots, y_{t}^{i}\}}$ be denoted as a set of ratings for each item $\smash{v^{i} \in \mathrm{X}_i}$. In this work, our goal is to correctly predict the rating score $\smash{y^{i}_{t+1}}$ of the next-item $\smash{v^{i}_{t+1}}$ based on the past sequences $\smash{\mathrm{X}_i}$ and $\smash{\mathrm{Y}_i}$, and we formulate this problem as a regression task. 

\subsection{Meta-Learning}
MAML \cite{finn2017model} is a gradient-based meta-learning algorithm for fast task adaptation, particularly few-shot learning problems. It can be easily integrated with any neural networks optimized by gradient descent. In this work, we assume that each user $u$ is a task $\mathcal{T}_i$ drawn from a task distribution $\mathnormal{p}(\mathcal{T})$. Each task $\mathcal{T}_i$ consists of two disjoint sets; support set and query set. The support set $\smash{\mathcal{D}^S_i = \{\mathrm{X}^S_i, \mathrm{Y}^S_i\}}$ is used for inner-loop optimization (task / user adaptation), and the query set $\smash{\mathcal{D}^Q_i = \{\mathrm{X}^Q_i, \mathrm{Y}^Q_i\}}$ is used for outer-loop optimization (meta-optimization). Based on this setting, our goal is to meta-learn desirable initial parameters of a given sequential recommender network that make the recommender quickly adapt to cold-start new users with a limited number of user-item interactions.

\begin{figure*}[htp]
    \centering
    \includegraphics[width=\textwidth]
    {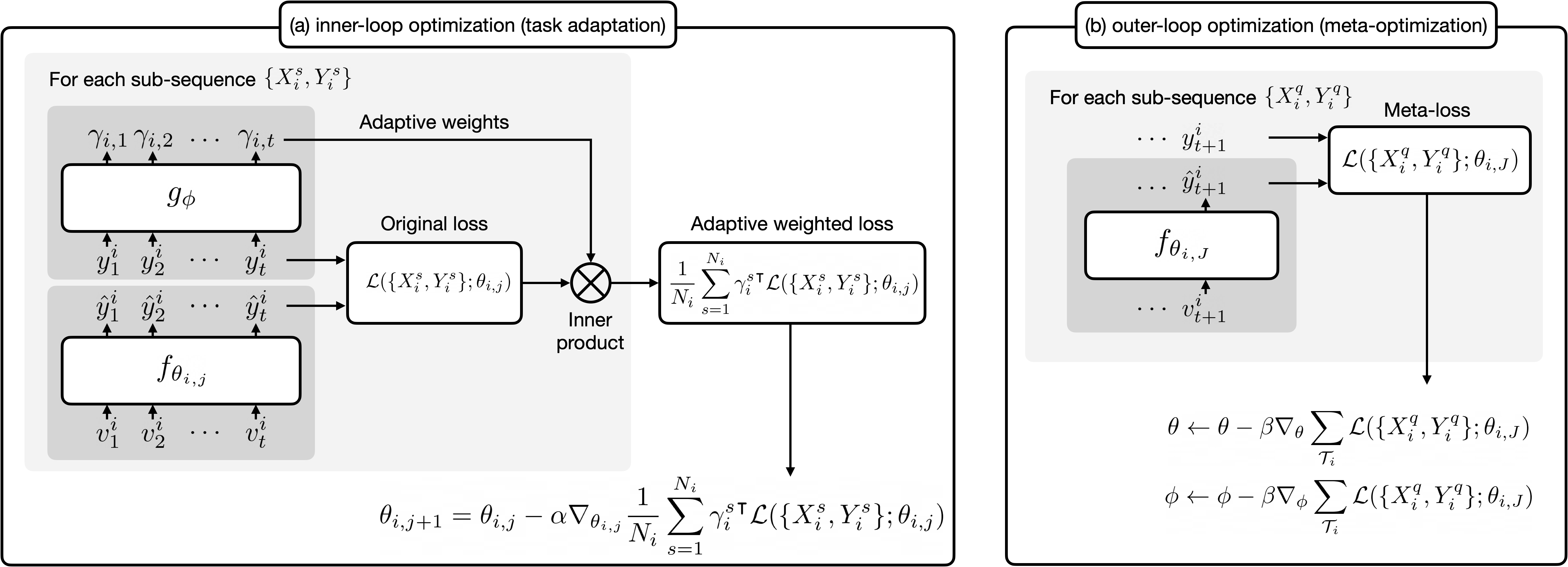}
    \caption{\textbf{Overview of MELO framework.}
    (a) The sequential recommender $f_{\theta_{i,j}}$ receives item sequences and predicts ratings. The task state recurrent encoder $g_{\phi_i}$ receives rating sequences and outputs adaptive weights. Each original loss is multiplied by each weight to calculate the adaptive weighted loss. Then, it locally updates $\theta_{i,j}$ for $J$ times.
    (b) Both $\theta$ and $\phi$ are evaluated by the query set and globally updated in the outer-loop.}
    \label{fig:2}
\end{figure*}

\section{Methodology}
The overall framework is detailed in Algorithm~\ref{alg:1} and Figure~\ref{fig:2}. We initially introduce a meta-learner, designed to adapt sequential recommenders to cold-start users. We demonstrate its ease of integration with any neural network-based sequential recommenders. Subsequently, we propose our adaptive weighted loss, which dynamically calibrates the learning process in response to imbalanced rating distributions.

\subsection{Meta-Learner}
Let us assume that we have a sequential recommender network $f_\theta$, and our goal is to learn a proper parameter initialization $\theta (=\theta_{i,0})$ that allows the network  $f_\theta$ to make fast adaptation to cold-start users with a few numbers of user-item interactions. We make this happen by introducing a meta-learner based on the MAML algorithm, which can be easily integrated with any neural network model optimized by gradient descent. The MAML algorithm performs bi-level optimization, which consists of local update (task / user adaptation) and global update (meta-optimization). We define the learnable initialization $\theta (=\theta_{i,0})$ as meta-knowledge, and it is updated to task-wise (each user) model parameters $\theta_{i,j}$ for each task $\mathcal{T}_i$ during the local update. The local update is also known as the inner-loop optimization process in that we conduct  $j$-step gradient updates with the training (support) dataset $\mathcal{D}^S_i$ as:
\begin{equation}
    \label{eqn:2}
    \theta_{i,j} = \theta_{i,j-1} - \alpha\nabla_{\theta_{i,j-1}}\mathcal{L}(\mathcal{D}^S_i;\theta_{i,j-1}),
\end{equation}
where $\alpha$ is the step size for the inner loop gradient-based optimization.

Each task-specific model $f_{\theta_{i,J}}$ is obtainable through $J$-step of gradient updates, and then it is evaluated with unseen query examples $\smash{\mathcal{D}^Q_i}$ for the task $\mathcal{T}_i$ in the outer-loop to properly find the generalized initialization $\theta$ (meta-parameter). We then express the outer-loop optimization process as:
\begin{equation}
    \theta \leftarrow \theta - \beta\nabla_{\theta}\sum_{\mathcal{T}_i}\mathcal{L}(\mathcal{D}^Q_i;\theta_{i,J}),
\end{equation}
where we aggregate multiple task-wise query losses to compute the gradient with respect to the initialization $\theta$, and $\beta$ is the step size for the outer loop gradient-based optimization.

To further improve generalization in meta-learning, we correctly split the overall user-item interactions (user-level sequence data) for task adaptation and meta-optimization, which correspond to support and query sets, respectively. For each user $u_i$, we hold out the next interaction $\smash{v_{t+1}^{i}}$ at time step $t+1$ and assign a sequence $\smash{\{v_{1}^{i}, v_{2}^{i}, \dots, v_{t+1}^{i}\}}$, which includes it as the last item, to the query set $\mathcal{D}^Q_i$ to evaluate the task-specific parameter $\theta_{i, J}$ in the outer-loop. 
On the other hand, we differently allocate the other sequence $\smash{\{v_{1}^{i}, v_{2}^{i}, \dots, v_{t}^{i}\}}$, which ends before the held-out interaction at time step $t+1$, to the support set $\mathcal{D}^S_i$, and this is mainly used to fine-tune the model in the inner-loop. Each example in both $\smash{\mathcal{D}^S_i}$ and $\smash{\mathcal{D}^Q_i}$ is generated using the data augmentation technique of previous studies \cite{huang2022learning, liu2018stamp, tan2016improved}. For example, let the assigned full sequence of items in the support set be $\smash{\mathrm{X}_{i}^{S}}$, then it is sliced into $N_i$ consecutive sub-sequences $\smash{\mathrm{X}_{i}^{s} \in \mathrm{X}_{i}^{S}}$: $\smash{\{v_{1}^{i}, v_{2}^{i} \}}$, $\smash{\{v_{1}^{i}, v_{2}^{i}, v_{3}^{i}\}}$, $\dots$, $\smash{\{v_{1}^{i}, v_{2}^{i}, v_{3}^{i}, \dots, v_{t}^{i}\}}$. On the contrary, we apply the slicing technique in reverse order to the query set $\smash{\mathrm{X}_{i}^{Q}}$ to generalize meta-knowledge $\theta$ based on the held-out next-item: $\{v_{t}^{i}, v_{t+1}^{i} \}$, $\{v_{t-1}^{i}, v_{t}^{i}, v_{t+1}^{i} \}$, $\dots$, $\{v_{1}^{i}, \dots, v_{t-1}^{i}, v_{t}^{i}, v_{t+1}^{i}\}$.

\subsection{Adaptive Weighted Loss}
In general, MAML assumes that the number of instances per class does not vary (i.e., $N$-way $k$-shot setting). This setting limits the model to properly utilize the learned meta-knowledge when the number of instances is disproportionate in a given task. Therefore, in this work, we aim to build sequential recommender systems that can adapt to cold-start users by incorporating the MAML algorithm and further rightly understand the state of imbalance distribution in each cold-start user and appropriately modify the learning process to prevent overfitting and achieve a good generalization. To do so, it is crucial to ensure that the sample-wise gradients are well-aligned without being overwhelmed by each other~\cite{yu2020grad}. We implement this by assigning an adaptive weight (gradient step size) to each sub-sequence loss.

To adaptively correct the learning path of imbalanced problems with the MAML-based sequential recommender systems, we employ the task state recurrent encoder that can effectively aggregate the state of rating distribution based on temporal dynamics in each task. Our proposed recurrent encoder $g_{\phi}$ meta-learns representation of the rating sub-sequence $\smash{\mathrm{Y}^{s}_i} \in \smash{\mathrm{Y}^{S}_i}$ by updating a recurrent hidden state vector $\smash{h^{i}_t}$ and further outputs the adaptive weight values $\gamma_i^{s} \in \mathbb{R}_{\geq 0}^{|\mathrm{Y}^{s}_i|}$ to correct the original inner-loop loss (task / user adaptation). For each item $v_t^i$ in the sub-sequence $\mathrm{Y}^{s}_i$, the recurrent cell $\smash{\hat{g}_{\phi^{\text{cell}}}}$ updates the hidden state vector $h^{i}_{t}$ from the previous ones with its rating score $y^{i}_t$, and the adaptive weight $\gamma_{i, t}$ is computed by transforming the hidden state vector as:
\begin{equation}
    \gamma_{i, t} = \tilde{g}_{\phi^{\text{out}}}(h^{i}_t) \in \mathbb{R}_{\geq 0}, \quad h^{i}_t = \hat{g}_{\phi^{\text{cell}}}(y^{i}_t; h^{i}_{t-1}),
\end{equation}
where the function $\tilde{g}_{\phi^{\text{out}}}$ is a linear mapping, which aligns item-wise gradients without being crushed by each other, and we compute the adaptive weights $\gamma_i^{s} = \left[ \gamma_{i, 1}, \gamma_{i, 2}, \dots,  \gamma_{i, |\mathrm{Y}^{s}_i|}\right]^{T}$ for all items in the sub-sequence $\mathrm{Y}^{s}_i$. This overall process can be expressed as $\gamma_i^{s} = g_{\phi}(\mathrm{Y}^{s}_i)$, and the encoder $g_\phi$ is only updated in the outer-loop as meta-knowledge that is shared over all tasks without any adaptation.

The original adaptation loss of each sub-sequence $\{\mathrm{X}^{s}_i, \mathrm{Y}^{s}_i\}$ is defined by $\mathcal{L}(\{\mathrm{X}^{s}_i, \mathrm{Y}^{s}_i\}; \theta_{i,j})$, and it is a vector composed of item-wise loss values. We correct the original adaptation loss by adaptively weighting item-wise loss values and obtain the modified loss to adapt appropriately to imbalanced problems. As both the adaptive weights $\gamma_i^{s}$ and the original adaptation loss values $\mathcal{L}(\{\mathrm{X}^{s}_i, \mathrm{Y}^{s}_i\}; \theta_{i,j})$ are equally defined over given items in the sub-sequence $\mathrm{X}^{s}_i$ (and $\mathrm{Y}^{s}_i$), we do inner product between them to aggregate the modified loss values and repeat this over all existing $N_i$ sub-sequences. That way, we accordingly run each inner-loop update as:
\begin{equation}
\theta_{i,j+1} = \theta_{i,j} - \alpha\nabla_{\theta_{i,j}}\frac{1}{N_i}\sum_{s=1}^{N_i}{\gamma^s_i}^{\intercal}\mathcal{L}(\{\mathrm{X}^{s}_i, \mathrm{Y}^{s}_i\};\theta_{i,j})
\end{equation}

\begin{algorithm}[t]
\caption{MELO}
\label{alg:1}
\begin{flushleft}
\textbf{Require: } $\mathnormal{p}(\mathcal{T})$: distribution over tasks \\
\textbf{Require: } $\alpha$, $\beta$: step size hyperparameters \\
\textbf{Require: } Sequential recommender network $\mathnormal{f_\theta}$ \\
\textbf{Require: } Task state encoder network $\mathnormal{g_\phi}$ \\
\begin{algorithmic}[1]
\STATE Randomly initialize $\theta$, $\phi$
 \WHILE{not done}
  \STATE Sample a batch of tasks $\mathcal{T}_i \sim \mathnormal{p}(\mathcal{T})$
  \FOR{all $\mathcal{T}_i$}
   \STATE Sample $\{\mathrm{X}^S_i, \mathrm{Y}^S_i\}$ sequences as support set $\mathcal{D}^S_i$ and $\{\mathrm{X}^Q_i, \mathrm{Y}^Q_i\}$ sequences as query set $\mathcal{D}^Q_i$ from $\mathcal{T}_i$
   \STATE Initialize adapted parameters $\theta_{i,0} = \theta$
   \FOR{inner-loop updates $j=0:\mathnormal{J}-1$}
    \STATE Compute the original inner loss on each sub-sequence: $\{\mathcal{L}(\{\mathrm{X}^s_i, \mathrm{Y}^s_i\};\theta_{i,j})\}_{s=1}^{N_i}$
    \STATE Compute the adaptive weight on each rating sequence: $\{\gamma_i^s = g_{\phi}(\mathrm{Y}^s_i)\}^{N_i}_{s=1} $
    \STATE Compute the adaptive weighted loss: \\
    $\tilde{\mathcal{L}}(\mathcal{D}^S_i;\theta_{i,j})=\frac{1}{N_i}\sum_{s=1}^{N_i}{\gamma_i^s}^\intercal \mathcal{L}(\{\mathrm{X}^s_i, \mathrm{Y}^s_i\};\theta_{i,j})$
    \STATE update $\theta_{i,j}$ for task adaptation: \\
    $\theta_{i,j+1} = \theta_{i,j} - \alpha\nabla_{\theta_{i,j}}\tilde{\mathcal{L}}(\mathcal{D}^S_i;\theta_{i,j})$
   \ENDFOR
   \STATE Compute the outer loss on the query set: \\ $\mathcal{L}(\mathcal{D}^Q_i;\theta_{i,J})$
  \ENDFOR
  \STATE update $(\theta,\phi)$ for meta-optimization: \\
  $\theta \leftarrow \theta - \beta\nabla_{\theta}\sum_{\mathcal{T}_i}\mathcal{L}(\mathcal{D}^Q_i;\theta_{i,J})$
  $\phi \leftarrow \phi - \beta\nabla_{\phi}\sum_{\mathcal{T}_i}\mathcal{L}(\mathcal{D}^Q_i;\theta_{i,J})$
\ENDWHILE
\end{algorithmic}
\end{flushleft}
\end{algorithm}

\begin{table}[t]
\centering
\caption{Dataset statistics after preprocessing}
{\tiny
\renewcommand{\arraystretch}{1.0}
\resizebox{1.00\columnwidth}{!}{
\begin{tabular}{lrrrr}
\toprule[0.7pt]
 & \textbf{Grocery} & \textbf{Sports} & \textbf{Yelp} & \textbf{Movie} \\[0.5ex] \hline
\#Users & 58,035 & 67,523 & 41,365 & 6,040  \\
\#Items & 4,776 & 13,448 & 1,221 & 2,514  \\
\#Ratings & 475,247 & 721,498 & 445,070 & 977,839 \\
Average length$^1$ & 8.0 & 10.7 & 10.8 & 29.2 \\
Balance score & 0.57 & 0.60 & 0.85 & 0.90 \\[-0.5ex]
\bottomrule[0.7pt]
\end{tabular}
}}
    \flushleft
    \footnotesize
    $^1$ The maximum length of a user-item sequence is 30.
\label{table:1}
\end{table}

\subsection{Encoder Architecture}
The role of the task state recurrent encoder is to interpret the rating distribution and convey valuable information to the recommender network, which learns the interaction between items and ratings. One natural way to model the user rating information is to use hand-designed features. MeTAL \cite{baik2021meta} preprocesses task information for its inputs (e.g., the mean of predictions and labels) and employs a 2-layer MLP that returns a scalar value as output. Bayesian TAML \cite{lee2019learning} uses a bayesian inference network using the statistics pooling technique with the mean, variance, and cardinality. However, this bayesian inference network has a complex structure and sometimes incurs unnecessary computation.

Our proposed encoder sequentially receives user-item ratings over multiple time steps and refines them into an informative representation without relying on preprocessed statistics and complex networks. When the encoder takes a user-item sequence, it measures the importance of each rating based on the current state. Then, it produces item-wise weights to adaptively balance the original loss value during the inner-loop updates. The encoder's recurrent structure determines which items to prioritize based on temporal dynamics, obviating the need for supplementary preprocessed data. Additionally, our encoder is not bound by any constraints, so it is easily adapted to various recommendation scenarios. A comparative analysis of different task state encoders is detailed in the experimental section 5.3.

\section{Experiments}
We conduct comprehensive experiments to evaluate the proposed method on four real-world datasets. Our goal is to address the following research questions (RQs):

\begin{itemize}
    \item \textbf{RQ1}: How much does MELO improve the performance of state-of-the-art sequential recommenders? (Section 5.2)
    \item \textbf{RQ2}: How do different loss functions affect the cold-start recommendation performance? (Section 5.3)
    \item \textbf{RQ3}: How robust is MELO to the cold-start problem with extremely short sequences? (Section 5.4)
    \item \textbf{RQ4}: How much does MELO reduce the high computational cost of the MAML algorithm? (Section 5.5)
    \item \textbf{RQ5}: How does each component of MELO affect the framework? (Section 5.6)
    \item \textbf{RQ6}: How well does MELO predict the ratings of different user types? (Section 5.7)    
\end{itemize}

\subsection{Experimental Setup}
\subsubsection{\textbf{Datasets.}} In this work, we use four public datasets containing users' 1-5 scale rating information on items. Each dataset differs in size and degree of rating imbalance. We assume a cold-start situation with a lack of user information, so we mainly select a small-sized dataset with fewer than 1 million records.

\begin{itemize}
    \item \textbf{Grocery}: 
    This dataset contains the ``Grocery'' category of product reviews from Amazon.com~\cite{ni2019justifying}. It has the most disproportionate ratings and the shortest sequence length. The proportion of each rating score is as follows: 5 (\textbf{72.8\%}), 4 (\underline{13.4\%}), 3 (6.7\%), 2 (3.8\%), 1 (3.3\%).
    \item \textbf{Sports}: 
    This dataset comprises the ``Sports'' category of product reviews from Amazon.com~\cite{ni2019justifying}. It has more users, items, and ratings than the Grocery category. The proportion of each rating score is as follows: 5 (\textbf{69.4\%}), 4 (\underline{18.0\%}), 3 (7.0\%), 2 (2.9\%), 1 (2.7\%).
    \item \textbf{Yelp}: 
    This dataset includes users, businesses, and reviews from the Yelp dataset challenge. The proportion of each rating score is as follows: 5 (\textbf{35.5\%}), 4 (\underline{35.2\%}), 3 (17.2\%), 2 (7.8\%), 1 (4.3\%).
    \item \textbf{Movie}: 
    GroupLens Research collected movie reviews from the MovieLens service. We adopt ``MovieLens 1M'' dataset. The proportion of each rating score is as follows: 5 (\textbf{36.2\%}), 4 (\underline{31.7\%}), 3 (16.0\%), 2 (8.8\%), 1 (7.3\%).
\end{itemize}

We preprocess the dataset to set a user cold-start scenario: (1) Additional information, such as age and category, is removed. Only user id, item id, rating, and timestamp information are left, representing cold-start users and their imbalanced rating distribution. (2) Items with fewer than 50 ratings are removed to avoid the item cold-start problem, which is not the primary focus of this study. (3) We build a user-item interaction sequence in chronological order and remove the timestamp feature afterward.

In quantifying the degree of imbalance, we use the Shannon entropy $H$ to measure the balance of rating distribution. For example, let a dataset have $n$ instances and $k$ classes of size $c_i$, then $H$ is calculated as $\log{k}$ when the size of all classes are equal $\frac{n}{k}$. Therefore, we can express the \textit{Balance} score as follows:
\begin{equation}
    \textit{Balance} = \frac{H}{\log{k}} = \frac{-\Sigma^k_{i=1}\frac{c_i}{n}\log{\frac{c_i}{n}}}{\log{k}}
\end{equation}
The \textit{Balance} score is closer to 0 when the dataset is imbalanced and equal to 1 when the dataset is entirely balanced. For example, the Grocery dataset is highly skewed, and its balance score is 0.57. In Table \ref{table:1}, we present the statistics of each preprocessed dataset.

\subsubsection{\textbf{Evaluation Metrics.}} 
To evaluate the experimental results for the proposed method, we adopt Root Mean Squared Error (RMSE) and Mean Absolute Error (MAE), which are standard metrics that address how close the predicted ratings are to the actual user ratings \cite{chai2014root}. The two metrics, RMSE and MAE, are complementary. RMSE disproportionately penalizes significant errors by squaring the residuals. As a result, it is more affected by wrong predictions. On the other hand, MAE weighs large and small errors equally. Therefore, it is preferred when outliers do not severely impact the objective.

Since the imbalanced distribution causes a large variance, we conduct performance evaluations focusing on RMSE. If we predict with one single score that accounts for the most, MAE can be pretty low, but RMSE is not. For example, MAE is 0.5140 if we predict all outcomes as a score of 5 in the Grocery dataset, which is lower than the score predicted by the proposed method. However, RMSE will be 1.1278 due to being penalized by a significant deviation.

\subsubsection{\textbf{Baselines.}} 
We reproduce the following representative sequential recommenders for a rating prediction task to verify the effectiveness of the proposed method.
\begin{itemize}
    \item \textbf{GRU4Rec} 
     \cite{hidasi2015session}: It uses Gated Recurrent Unit (GRU) to encode user-item interaction records into a representation vector for the session-based recommendation. The original GRU4Rec is used with an additional linear projection layer to produce a rating value.
    \item \textbf{NARM} 
     \cite{li2017neural}: It integrates the GRU-based local and global encoder with the attention mechanism to capture the user's primary purpose in the current session. Instead of concatenating local and global contexts as in the original paper, we multiply them and add a linear projection layer. 
    \item \textbf{SASRec} 
     \cite{kang2018self}: It employs the unidirectional transformer-based method with single-head attention for the sequential recommendation. The original SASRec is used with an additional linear projection layer to produce a rating value.
    \item \textbf{BERT4Rec} 
     \cite{sun2019bert4rec}: It uses the bidirectional multi-head self-attention architecture to effectively model sequential patterns in user-item interaction. The original BERT4Rec is used with an additional linear projection layer to produce a rating value.
\end{itemize}

\subsubsection{\textbf{Implementation Details.}} 
Sequential recommender baselines are implemented based on the codes provided by the original authors. We use simple dictionary embedding with embedding matrix $W\in\mathbb{R}^{m\times d_v }$ for all baseline models, where $m$ denotes the number of items and $d_v$ denotes the dimension size of embedding. Positional embedding is added for SASRec and BERT4Rec.

Our meta-learner is implemented with PyTorch referring to MAML \cite{finn2017model} and MAML++ \cite{antoniou2018train}. We divide users into non-overlapping partitions for training, validation, and test sets. The maximum user-item sequence length used in this experiment is 30 to create cold-start situations. The user sequence selected as the task is split into 25 examples of support set and 3 examples of query set. We use MSE for the original loss function, and the number of inner-loop steps $J$ is 3 in default. We use LSTM cells~\cite{hochreiter1997long} for our recurrent encoder, which is composed of rating embedding with a dimension size of 16 and a hidden state dimension of 32. We normalize all ratings between 0 and 1, and clip the gradients to prevent the gradient explosion. A cosine annealing schedule is applied along with the Adam \cite{kingma2014adam} optimizer. We train 32,000 episodes for the Movie dataset and 48,000 for the Grocery, Sports, and yelp datasets. For evaluation, we use 600 episodes for validation and 1,000 episodes for the test. We perform 5 independent experiments for all baselines and report the average value. Our code contains all the details on every baseline architecture and hyperparameter.\footnote{\url{https://github.com/YangYongJin/MELO}}

\begin{table*}[ht]
\centering
\caption{Performance comparison of sequential recommender methods with or without MAML and MELO integration. \textbf{Bold} represents the best variant in each sequential recommender, and \underline{underlined} indicates the second best variant. Note that * denotes improvement over the second-best variant with p-value $<$ 0.05 (measured by t-test with 5 independent trials.)}
{\Large
\setlength{\tabcolsep}{4pt}
\renewcommand{\arraystretch}{1.8}
\resizebox{1.00\textwidth}{!}{
\begin{tabular}{lccllccccccccccccccclrr}
\toprule[1.5pt]
\multirow{2}{*}{\textbf{Dataset}} & \multicolumn{1}{c}{\multirow{2}{*}{\textbf{\begin{tabular}[c]{@{}c@{}}Avg.\\[-2.5ex]
length \end{tabular}}}} & \multirow{2}{*}{\textbf{\begin{tabular}[c]{@{}c@{}}Balance\\[-2.5ex] score \end{tabular}}} & \multirow{2}{*}{\textbf{Metric}} & & \multicolumn{3}{c}{\textbf{GRU4Rec}} &  & \multicolumn{3}{c}{\textbf{NARM}} &  & \multicolumn{3}{c}{\textbf{SASRec}} &  & \multicolumn{3}{c}{\textbf{BERT4Rec}} &  & \multicolumn{1}{c}{\multirow{2}{*}{\textbf{\begin{tabular}[c]{@{}c@{}}Improv.\\[-2.5ex]
over \\[-2.5ex]
Basic$^1$ \end{tabular}}}} & \multicolumn{1}{c}{\multirow{2}{*}{\textbf{\begin{tabular}[c]{@{}c@{}}Improv.\\[-2.5ex]
over \\[-2.5ex]
MAML$^2$ \end{tabular}}}} \\
\cline{6-8} \cline{10-12} \cline{14-16} \cline{18-20}
& & & & & \textbf{Basic} & \textbf{MAML} & \textbf{MELO} &  & \textbf{Basic} & \textbf{MAML} & \textbf{MELO} & & \textbf{Basic} & \textbf{MAML} & \textbf{MELO} & & \textbf{Basic} & \textbf{MAML} & \textbf{MELO} & & \multicolumn{1}{c}{} & \multicolumn{1}{c}{} \\ \hline
\multirow{2}{*}{\shortstack[c]{Grocery}}           
& \multirow{2}{*}{\shortstack[c]{8.0}} & \multirow{2}{*}{\shortstack[c]{0.57}} & RMSE & & 1.0794 & \underline{1.0293}  & $\textbf{0.9859}^*$ &  & 1.0804  & \underline{1.0176}  & $\textbf{0.9935}^*$ &  & 1.0911  & \underline{1.0651} & $\textbf{0.9949}^*$ &  & 1.0907 & \underline{1.0501} & $\textbf{0.9925}^*$ &  & 8.63\% & 4.67\%  \\
& & & MAE &  & 0.7797 & \underline{0.6890}  & $\textbf{0.6700}^*$ &  & 0.7927 & \underline{0.6811}  & $\textbf{0.6743}^*$ & & 0.8003 & \underline{0.7352} & $\textbf{0.6724}^*$ &  & 0.7971 & \underline{0.6803} & $\textbf{0.6608}^*$ &  & 15.53\% & 3.79\% \\ \hline

\multirow{2}{*}{\shortstack[c]{Sports}}
& \multirow{2}{*}{\shortstack[c]{10.7}} & \multirow{2}{*}{\shortstack[c]{0.60}} & RMSE & & 1.0364 & \underline{1.0147} & $\textbf{0.9710}^*$ & & 1.0373 & \underline{1.0007}  & $\textbf{0.9667}^*$ &  & 1.0360 & \underline{1.0134} & $\textbf{0.9690}^*$ &  & 1.0332 & \underline{1.0041}  & $\textbf{0.9565}^*$ &  & 6.75\% & 4.21\% \\
& & & MAE & & 0.7518 & \underline{0.6987} & $\textbf{0.6765}^*$ &  & 0.7837 & \underline{0.6887}  & $\textbf{0.6627}^*$ &  & 0.7558 & \underline{0.6695} & \textbf{0.6624} &  & 0.7656 & \underline{0.6820} & $\textbf{0.6464}^*$ &  & 13.34\% & 3.31\% \\ \hline

\multirow{2}{*}{\shortstack[c]{Yelp}}             
& \multirow{2}{*}{\shortstack[c]{10.8}} & \multirow{2}{*}{\shortstack[c]{0.85}} & RMSE
& & \underline{1.2084} & 1.2396 & \textbf{1.2005}
& & 1.2156 & \underline{1.2154} & $\textbf{1.2049}^*$
& & \underline{1.2374} & 1.2540 & $\textbf{1.1931}^*$
& & \underline{1.2422} & 1.2610 & $\textbf{1.1777}^*$
& & 2.58\% & 3.87\% \\
& & & MAE           
& & \underline{0.9765} & 0.9994 & $\textbf{0.9666}^*$
& & \underline{0.9719} & 0.9825 & \textbf{0.9688}
& & \underline{0.9841} & 1.0008 & $\textbf{0.9591}^*$
& & \underline{0.9919} & 1.0106 & $\textbf{0.9431}^*$
& & 2.20\% & 3.88\% \\
\hline

\multirow{2}{*}{\shortstack[c]{Movie}}        
& \multirow{2}{*}{\shortstack[c]{29.2}} & \multirow{2}{*}{\shortstack[c]{0.90}} & RMSE                             
& & 0.9937 & \underline{0.9856} & $\textbf{0.9725}^*$
& & 0.9939 & \underline{0.9818} & $\textbf{0.9733}^*$
& & 0.9944 & \underline{0.9761} & \textbf{0.9675}
& & 0.9990 & \underline{0.9755} & \textbf{0.9750}
& & 2.33\% & 0.78\% \\
& & & MAE
& & 0.7930 & \underline{0.7744} & \textbf{0.7674}
& & 0.7943 & \textbf{0.7666} & \underline{0.7702}
& & 0.7929 & \underline{0.7642} & \textbf{0.7602}
& & 0.7991 & \textbf{0.7617} & \underline{0.7685}
& & 3.55\% & 0.02\% \\ 
\bottomrule[1.5pt]
\end{tabular}
}}
    \flushleft
    \footnotesize
    $^1$ The average improvement over Basic baselines \\
    $^2$ The average improvement over MAML integrations
\label{table:2}
\end{table*}

\subsection{Overall Results (RQ1)}
Table \ref{table:2} summarizes the experimental results for all baseline methods: Basic, MAML, and MELO integration. MELO effectively integrates state-of-the-art sequential recommender networks, and the average performance of MELO integrations surpasses both Basics and MAML integrations across all datasets. Specifically, the most significant performance improvement is observed when the dataset exhibits a severe imbalance and a short sequence length, as seen in the Grocery, Sports, and Yelp datasets. Notably, MELO shows the greatest performance improvement compared to MAML on the Grocery dataset. This dataset demonstrates the most extreme imbalance with 72.8\% of ratings clustered at a score of 5 and an average sequence length of 8.

In contrast, on the Movie dataset, which has an average sequence length of 29.2 (considered relatively long), and a balance score of 0.9 (indicating relatively balanced), the performance improvement between the existing MAML algorithm and MELO is not significant. This suggests that the effect of adaptive loss is pronounced when user information is limited. Further examination of the Movie dataset to assess the performance impact of different sequence lengths will be conducted in section 5.4.

While MELO surpasses both MAML and Basic on the Yelp dataset, MAML's performance is inconsistent, often underperforming when compared to Basic. Although both MELO and MAML utilize the gradient-based meta-learning algorithm, their performance varies depending on whether adaptive loss is applied. From this result, we can infer that the proposed adaptive weighting strategy enhances the robustness of the gradient-based meta-learning method in various real-world situations.

Overall, the experimental results suggest that the contribution of the adaptive loss becomes more significant when faced with limited and imbalanced data. Given that MELO consistently outperforms other methods in most cases, it can be considered a standard solution for various user cold-start scenarios.

\subsection{Loss Function (RQ2)}
To better understand the role of the loss function, we conduct a comparative experiment with three different ways of generating the adaptive loss function: Focal, Stats, and Ours (adaptive weighted loss). We use BERT4Rec as a base recommender for this experiment. Table~\ref{table:3} displays the performance of three different loss functions.

In this experiment, we transform the Focal loss equation for a regression task (a rating prediction task) with the regularized MSE loss based on the previous study \cite{lu2018deep}. The Focal loss \cite{lin2017focal} decreases the relative loss for easy and correct examples but focuses on learning hard and incorrect examples. The scaling factor of the Focal loss downsizes the contribution of easy examples by dynamically reducing cross-entropy loss during training. However, baselines applied with the Focal loss do not improve the performance but rather deteriorate. This result implies that a fixed re-weighting strategy by counting the number of easy and hard examples in the dataset cannot achieve user-level optimization.

Next, we conduct a comparative experiment with the Stats loss, which processes statistical features from the user-item sequence and provides further information to the recommender model. The Stats encoder is built with a 2-layer MLP referring to the previous study \cite{baik2021meta}. The statistical information of the target sequence, such as mean, variance, and cardinality, is preprocessed separately and put into the stats encoder network as input. Since each dataset has its cold-start characteristic, the preprocessing process becomes complex and domain knowledge dependent. Table~\ref{table:4} shows the best-performing combinations on the datasets.

Our adaptive weighted loss achieves comparable performance to the Stats loss. This result demonstrates that the representation modeled by our recurrent encoder has sufficient information compared to manually designed statistics. It allows us to compute adaptive weights more concisely without increasing the complexity of the network and aggregating additional information. Furthermore, our encoder models the user's rating sequentially, and the representation can be updated efficiently as new user information continuously arrives (e.g., online services). However, the Stats encoder requires a batch of samples to extract user information properly. In summary, individual user behavior modeling is essential and beneficial for calculating adaptive loss to improve personalized recommendation accuracy. Our adaptive weighted loss shows generalized performance over multiple users and datasets.

\begin{table}[t]
\centering
\caption{Impact of different loss functions}
{\tiny
\renewcommand{\arraystretch}{1.5}
\resizebox{0.9\columnwidth}{!}{
\begin{tabular}{lllccc}
\toprule[0.7pt]
\multicolumn{1}{c}{\textbf{Dataset}} & \multicolumn{1}{l}{\textbf{Metric}} &  & \textbf{Focal} & \textbf{Stats$^1$} & \textbf{Ours} \\[0.8ex] \hline
\multirow{2}{*}{Grocery} 
& RMSE &  & 1.0390 & \underline{0.9927} & \textbf{0.9925} \\
& MAE  &  & 0.6918 & \underline{0.6611} & \textbf{0.6608}  \\
\hline
\multirow{2}{*}{Sports}              
& RMSE &  & 1.0171 & \underline{0.9569} & \textbf{0.9565} \\
& MAE  &  & 0.7054 & \textbf{0.6326} & \underline{0.6464} \\
\hline
\multirow{2}{*}{Yelp}               
& RMSE &  & 1.2586 & \underline{1.1781} & \textbf{1.1777} \\
& MAE  &  & 1.0238 & \underline{0.9438} & \textbf{0.9431} \\ \hline
\multirow{2}{*}{Movie}           
& RMSE &  & 0.9833 & \textbf{0.9746} & \underline{0.9750} \\
& MAE  &  & 0.7700 & \textbf{0.7663} & \underline{0.7685} \\[-0.5ex]
\bottomrule[0.7pt]
\end{tabular}
}}
    \flushleft
    \footnotesize
    $^1$ The result is from the best-performing combination of Table~\ref{table:4}.\\
\label{table:3}
\end{table}

\begin{table}[t]
\centering
\caption{Performance of the Stats loss on different combinations of input features. Bold represents the best result on each dataset.}
{\tiny
\renewcommand{\arraystretch}{1.5}
\resizebox{1.0\columnwidth}{!}{
\begin{tabular}{llcccc}
\toprule[0.8pt]
\multicolumn{1}{l}{\textbf{Feature}} & \multicolumn{1}{l}{\textbf{Metric}} & \textbf{Grocery} & \textbf{Sports} & \textbf{Yelp} & \textbf{Movie} \\[0.6ex] \hline
(1) All features
& RMSE & 0.9940 & 0.9704 & 1.1881 & 0.9785 \\ \hline
(2) No mean
& RMSE & 1.0076 & 0.9649 & \textbf{1.1781} & 0.9840 \\ \hline
(3) No std
& RMSE & \textbf{0.9927} & 0.9709 & 1.1926 & \textbf{0.9746} \\
\hline
(4) No label$^1$
& RMSE & 1.0050 & 0.9661 &	1.1911 & 0.9817 \\
\hline
(5) No pred$^2$
& RMSE & 0.9984 & \textbf{0.9569} & 1.1844 & 0.9773 \\
\hline
(6) No loss$^3$
& RMSE & 0.9959 & 0.9663 & 1.1950 & 0.9753 \\
\hline
(2) + (3)
& RMSE & 1.0336 & 0.9610 & 1.2005 & 0.9765 \\
\hline
(4) + (5) + (6)
& RMSE & 1.0488 & 0.9868 & 1.1968 &	0.9871 \\[-0.6ex]
\bottomrule[0.8pt]
\end{tabular}
}}
    \flushleft
    \footnotesize
    $^1$ The true rating distribution of the sub-sequence $\mathrm{Y}^{s}_i$ \\
    $^2$ The predicted rating distribution of the sub-sequence $\mathrm{Y}^{s}_i$ \\
    $^3$ The original loss value computed by the sequential recommender network
\label{table:4}
\end{table}

\subsection{Severe Cold-Start Problem (RQ3)}
We investigate the effectiveness of MELO under severe cold-start conditions, focusing on the Movie dataset with its longest average sequence length of 29.2 (MELO can handle a maximum sequence length of 30). In this experiment, we randomly slice a sequence length from 5 to $T$ to generate diverse cold-start situations. As $T$ decreases, the percentage of severe cold-start users increases.

Table~\ref{table:5} illustrates the prediction performance of BERT4Rec with MELO and MAML integration on different maximum lengths $T$. The result shows that the performance improvement of MELO over MAML integration increases as the maximum length $T$ decreases. It implies that MELO is versatile under very challenging conditions, where the maximum sequence is only a length of 10, and all sequences lie between 5 and 10. In other words, short-length sequences are problematic for MAML integration alone. We may assume that adaptive loss affects the model to avoid over- and under-fitting with a small amount of interaction. This effect helps the recommender model to find a valid path to imbalanced cold-start users.

It is worth noting that the other datasets, namely Grocery, Sports, and Yelp, already have a high percentage of severe cold-start scenarios, given their average lengths are below 10.8. As shown in Table~\ref{table:2}, MELO performs well in these environments.

\begin{table}[t]
\centering
\caption{Performance with different maximum sequence length $T$ on the Movie dataset. \textbf{Bold} represents the best improvement by MELO among different sequence lengths, and \underline{underlined} indicates the second best improvement.}
{\scriptsize
\renewcommand{\arraystretch}{1.5}
\resizebox{1.00\columnwidth}{!}{
\begin{tabular}{llccccc}
\toprule[0.7pt]
 \textbf{Metric} & \textbf{Method} &  \textbf{10} & \textbf{15}     & \textbf{20}   & \textbf{25}     & \textbf{30}     \\[0.4ex] \hline
 \multirow{3}{*}{RMSE}
 & MAML & 0.9879 & 0.9802 & 0.9772 & 0.9806 & 0.9623 \\
 & MELO & 0.9629 & 0.9674 & 0.9599 & 0.9677 & 0.9581 \\ \cline{2-7} 
 & Improv. & \textbf{2.53\%} &	1.31\% & \underline{1.77\%} & 1.31\% & 0.44\% \\ \specialrule{0.7pt}{0pt}{0pt}
 \multirow{3}{*}{MAE}
 & MAML & 0.7772 & 0.7712 & 0.7649 & 0.7709 & 0.7545 \\
 & MELO & 0.7647 & 0.7671 & 0.7634 & 0.7688 & 0.7578 \\ \cline{2-7}
 & Improv. & \textbf{1.61\%} &	\underline{0.53\%} & 0.20\% & 0.27\% & -0.43\% \\                           
\bottomrule[0.7pt]
\end{tabular}
}}
\label{table:5}
\end{table}

\subsection{Model Efficiency (RQ4)}
\begin{figure}[t]
    \centering
    \includegraphics[width=1.00\columnwidth]{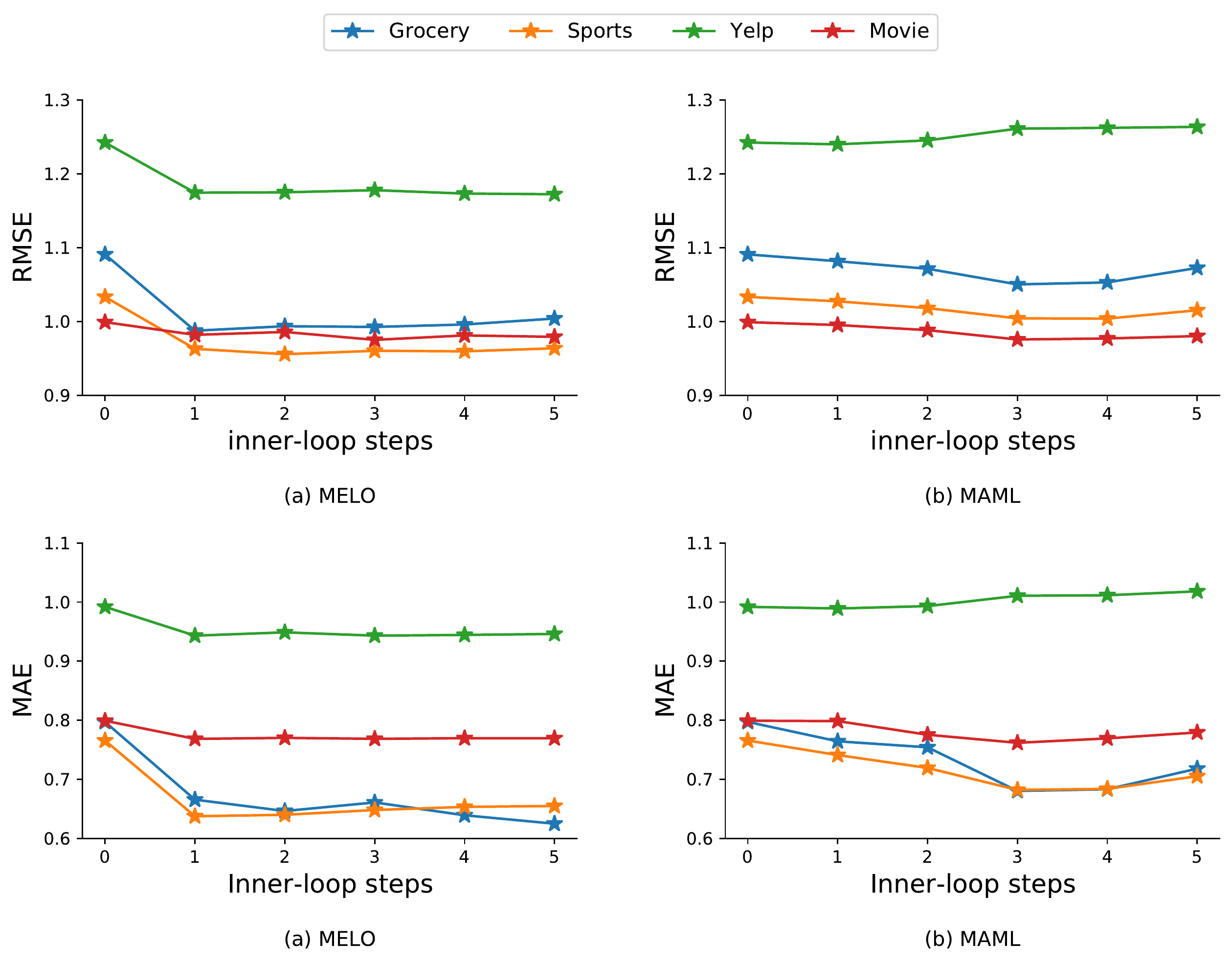}
    \caption{The performance of (a) MELO and (b) MAML according to the number of inner-loop steps. MELO reaches its optimal performance with one iteration.}
    \label{fig:3}
\end{figure}

We now examine the extent to which MELO reduces the inner-loop steps of the MAML algorithm. In this experiment, we utilize BERT4Rec as a sequential recommender network. Figure~\ref{fig:3} illustrates the performance of MELO and MAML as a function of the number of inner-loop updates on each dataset. Both the RMSE and MAE values noticeably decrease following a single update, with MELO achieving its optimal performance after only one iteration. This observation suggests that the adaptive loss function successfully propagates an adequate amount of loss to the recommender network during the task adaptation process, thereby reducing the number of necessary inner-loop optimization steps. Consequently, MELO guides the sequential recommender network to the optimal point more rapidly.

On the other hand, MAML requires more updates than MELO in order to achieve its peak performance. As illustrated in Figure~\ref{fig:3}, both the RMSE and MAE values decrease gradually in relation to the number of inner-loop steps to reach optimal performance. This trend aligns with the MAML's original paper, which also illustrated incremental performance improvement as the number of inner-loop updates increased~\cite{finn2017model}. Furthermore MAML's performance on the Yelp dataset indicates an overfitting problem from the first update, implying that merely increasing the number of updates does not improve performance when using the MAML algorithm alone.

In conclusion, the results show that MELO can achieve the optimal performance more quickly and cost-effectively than the traditional MAML methodology. This rapid adaptation capability enables MELO to overcome a fundamental limitation of the MAML algorithm, namely its requirement for extensive computations during the optimization process. This crucial characteristic enhances MELO's adaptability to real-world applications.

\subsection{Ablation Study (RQ5)}
We perform an ablation study over the critical components of MELO to investigate each contribution, including the sequential recommender (SR), the meta-learner (ML), and the adaptive loss (AL). Table~\ref{table:6} shows the results of MELO and its variants. We use BERT4Rec as a sequential recommender and conduct experiments on the Grocery dataset, which has the most imbalanced rating distribution. Hyperparameters of each variant are selected at their optimal settings.
\begin{enumerate}[(1)]
    \item \textbf{ML} and \textbf{AL}: We remove transformer layers from BERT4Rec and leave the output layer only. This result verifies that modeling sequential information is valuable for improving the cold-start recommendation performance.
    \item \textbf{SR} and \textbf{AL}: The result shows the advantage of the bi-level optimization from the meta-learning algorithm. It demonstrates the importance of inner-loop optimization and utilizing meta-knowledge for a personalized recommendation.
    \item \textbf{SR} and \textbf{ML}: This variant is the MAML-only integration without the benefit of adaptive loss. The MAML algorithm struggles to capture each user's state and improve the rating prediction performance.
    \item \textbf{SR}, \textbf{ML}, and \textbf{AL}: MELO, in which all components are combined, shows the best performance. The sequential recommender models a user's preferences, and the MAML algorithm identifies the optimal point of each user-specific learning with the support of the adaptive weights.
\end{enumerate}

\begin{table}[t]
\centering
\caption{Ablation study on the Grocery dataset}
{\small
\renewcommand{\arraystretch}{1.5}
\resizebox{1.00\columnwidth}{!}{
\begin{tabular}{c|p{3cm}|p{1.2cm}p{1.2cm}|rr}
\toprule[1pt]
  \textbf{\#} & \multicolumn{1}{c|}{\textbf{Architecture}} & \textbf{\hspace{2mm}RMSE} & \textbf{\hspace{3mm}MAE} & \multicolumn{2}{c}{\textbf{Improv.}} \\[0.8ex] \hline
(1) & ML + AL & \multicolumn{1}{c}{1.0583} & \multicolumn{1}{c|}{0.7235} & -6.21\% & -8.67\% \\
(2) & SR + AL & \multicolumn{1}{c}{1.0814} & \multicolumn{1}{c|}{0.7967} & -8.22\% & -17.06\% \\
(3) & SR + ML & \multicolumn{1}{c}{1.0501} & \multicolumn{1}{c|}{0.6803} & -5.49\% & -2.87\% \\
(4) & SR + ML + AL (Ours) & \multicolumn{1}{c}{\textbf{0.9925}} & \multicolumn{1}{c|}{\textbf{0.6608}} & - & - \\
\bottomrule[1pt]
\end{tabular}
}}
\label{table:6}
\end{table}

\subsection{Case Study (RQ6)}
In this section, we conduct a user case study to examine how MELO provides a personalized recommendation experience for individual users in an imbalanced environment. First, we select users who represent generous, fair, and grumpy preferences in the Grocery dataset, which displays the most imbalanced rating distribution. A generous user has a rating score of 1 (5\%), 2 (4\%), 3 (4\%), 4 (15\%), and 5 (72\%) proportion. A fair user's rating distribution is 1 (28\%), 2 (11\%), 3 (18\%), 4 (12\%), and 5 (31\%). A grumpy user has a rating score of 1 (58\%), 2 (17\%), 3 (10\%), 4 (7\%), and 5 (8\%) proportion. We then visualize the distribution of predicted ratings to compare how effectively the baseline methods account for rare ratings with few records.

\begin{figure}[t]
    \centering
    \includegraphics[width=0.9\columnwidth]{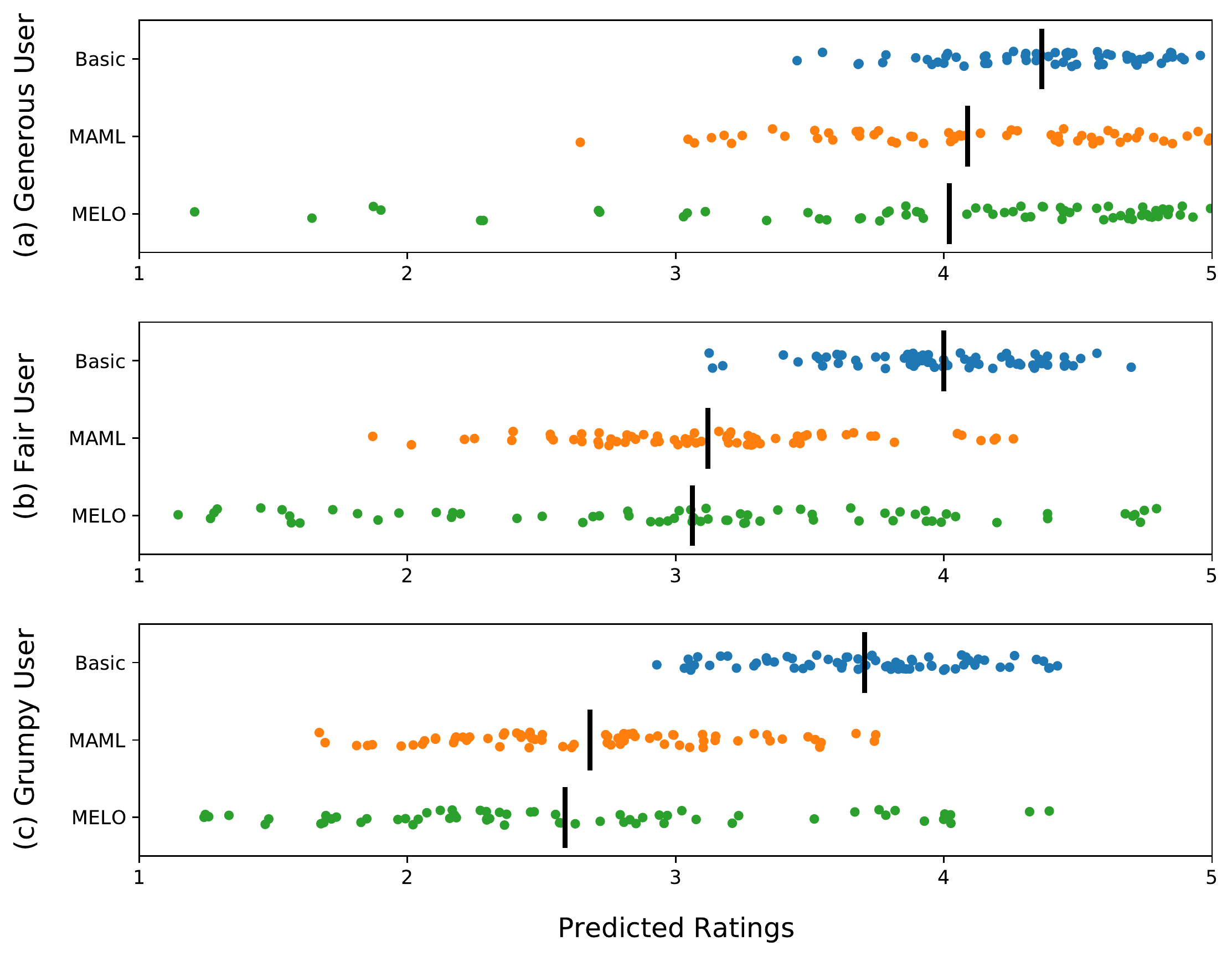}
    \caption{The predicted rating values for the representative user types: (a) generous, (b) fair, and (c) grumpy. The baseline methods predict a rating of time step $t+1$ by taking the sequence up to time step $t$ at each inference. The black bar represents the arithmetic mean of predicted values.}
    \label{fig:4}
\end{figure}

Figure~\ref{fig:4} shows the case study results. Basic methods fail to capture the preference and the degree of imbalance information from the user sequence. As a result, they bias their predictions towards the score with the most records. MAML displays better personalization performance but a slightly lower adaptation ability for imbalanced ratings. MELO shows a similar predictive average to MAML but offers a broader range of predictions, extending to ratings with few records. For example, while MAML fails to predict items rated as a score of 1 by the generous user, MELO achieves better prediction performance.

\subsection{Discussion}
While our approach presents a practical solution, there remains potential for future research. First, a deeper investigation into online recommender update strategies, especially within the framework of the proposed sequential recommender and task state recurrent encoder structure, could enhance the system's capability in adapting to users' up-to-date interests. Second, exploring and comparing alternative re-weighting strategies is imperative to identify the most effective ones for general recommendation algorithms. Such exploration could potentially enable the adaptive weighted loss approach to be applied in a wider range of scenarios, extending beyond recurrent relationships.

\section{Conclusion}
Our work is the first to tackle the problem of imbalanced ratings of cold-start users in the context of sequential recommender systems and meta-learning. In this study, we propose MELO, an approach that effectively captures the imbalanced rating distribution of cold-start users to provide personalized recommendations. Our extensive experiments on real-world datasets demonstrate that MELO outperforms the original MAML algorithm. This result underscores the importance of learning an adaptive weighted loss for each task, as opposed to employing a rule-based scaling strategy. Additionally, MELO's model-agnostic architecture allows easy integration with existing state-of-the-art sequential recommender networks, thereby enhancing performance. The task state recurrent encoder receives rating information sequentially and refines it into informative representations, producing adaptive weights without the need for manually designed statistics. In this way, we correct the original loss by learning the adaptive weighting strategy, which leads to a quicker adaptation to each cold-start user. We believe that our work for cold-start recommendation with imbalanced user feedback will serve as a momentous milestone for real-world applications.

\bibliographystyle{ACM-Reference-Format}
\balance
\bibliography{melo}

\end{document}